\documentclass[prd,twocolumn,nofootinbib]{revtex4}
\usepackage{latexsym}
\usepackage{epsfig}
\usepackage{amssymb}
\newcommand{\beq}{\begin{eqnarray}}
\newcommand{\eeq}{\end{eqnarray}}
\newcommand{\be}{\begin{eqnarray}}
\newcommand{\ee}{\end{eqnarray}}

\begin{document}

\title{Disformal quintessence}
\date{\today}
\author{Tomi S. Koivisto} 
\email{T.Koivisto@Thphys.Uni-Heidelberg.De}
\affiliation{Institute for Theoretical Physics, University of Heidelberg, 69120 Heidelberg,Germany}

\begin{abstract}

A canonic scalar field minimally coupled to a disformal metric generated by the field itself is considered. Causality and stability 
conditions are derived for such a field. Cosmological effects are studied and it is shown that the disformal 
modification could viably trigger an acceleration after a scaling matter era, thus possibly alleviating the 
coincidence problem. 

\end{abstract}

\maketitle

\section{Introduction}

Quantum effects in curved spacetime are known to introduce geometric corrections to the Einstein-Hilbert
action. Thus a de Sitter era for the early universe and the resolution of the initial singularity were predicted 
\cite{Starobinsky:1980te} from the semiclassical considerations of the spacetime itself. In the past few years, with 
the advent of the dark energy paradigm, the possibility has been contemplated that the present acceleration of the universe 
could also be derived from a more fundamental gravity theory superseding general relativity at 
large scales \cite{Nojiri:2003ft,Nojiri:2008nt}. This attractive prospect of unified inflation and dark energy (with perhaps 
dark matter) without ad hoc exotic matter sources has received a lot of attention \cite{Nojiri:2006ri,Nojiri:2008ku}. 
The prototype models are the nonlinear or $f(R)$ gravity among more general scalar-tensor theories \cite{Sotiriou:2008rp,Sotiriou:2007yd}. 
As well known \cite{Maeda:1988ab,Magnano:1993bd}, these models are related to coupled scalar field matter by conformal transformation 
\cite{Capozziello:2006dj,Bamba:2008hq}.    

One may ask whether relations of different forms could be considered as a consistent framework for gravity theories and 
their cosmological applications. It has been argued that a general Finslerian relation between matter and 
gravity geometries is restricted, by requring causality, validity weak equivalence principle and in particular covariance in
in its strictest sense, to a certain generalisation of the conformal relation \cite{Bekenstein:1992pj}. Let us therefore consider the class 
of the so called disformal transformations, which may depend on a scalar field $\phi$,
\be \label{disf}
\bar{g}_{\alpha\beta} = A(\phi)g_{\alpha\beta} + B(\phi) \phi_{,\alpha}\phi_{,\beta}. 
\ee
If $B=0$, the relation reduces to the conformal transformation which preserves the angles and the light cones. Any 
nonzero $B$ then causes a disformal modification which results in difference of the causal structures of the two
Riemannian geometries. This feature has been exploited to construct variable speed of light cosmology to 
cope without a usual inflation \cite{Clayton:1998hv,Clayton:2001rt}. On the other hand, an interestingly short inflation model with 
$A=-B=1$, has been considered in detail \cite{Kaloper:2003yf}. The disformal property is crucial in producing lensing phenomena 
in relativistic MOND models \cite{Bekenstein:2004ne,Skordis:2005xk}. It appears also within the so called Palatini 
formalism \cite{Koivisto:2005yc,Koivisto:2006ie} if the action includes Ricci tensor squared terms \cite{Allemandi:2004wn,Li:2007xw}. 

The main aim of the present paper is to further explore the cosmological possibilities of the relation 
(\ref{disf}) and in particular see how the disformal property could be used to unify dark energy within this 
framework. 
Previously, most authors couple matter to the disformal metric $\bar{g}_{\mu\nu}$, while 
gravity is given by the Einstein-Hilbert action of $g_{\mu\nu}$. The cosmological expansion is then effectively 
sourced by matter, whose energy density $\rho$ and the pressure $p$ are given by 
(dot denoting the time derivative)
\be \label{rho2}
\rho = \left(1-\frac{B}{A}\dot{\phi}^2\right)^{-\frac{1}{2}}\bar{\rho}, \quad p = 
\left(1-\frac{B}{A}\dot{\phi}^2\right)^{\frac{1}{2}}\bar{p}.
\ee
In usual coupled quintessence scenario \cite{Amendola:1999er}, the cosmon field $\phi$ \cite{Wetterich:1994bg} has
a conformal coupling to matter. The field should become energetically dominating recently to drive the acceleration. 
Now, we could consider a disformal extension of the usual 
coupled quintessence scenarios based on purely conformal (dilatation) symmetry considerations \cite{Wetterich:1987fm}.  
Then we would consider dark matter living in the metric disformed by a nonzero $B$. One notes that if the ''bare'' 
pressure vanishes, the coupling does not generate effective pressure. Then, as $B/A$ grows, the physically effective 
energy density just dilutes faster, and one does not expect new effects if the $\rho$ was not dynamically significant 
already earlier. Moreover, we have checked that viewing cosmology as a dynamical system, one does not find new fixed 
points by taking into account $B/A$: all fixed points require $B=0$. Thus a disformal coupling between dark 
matter and dark energy does not seem helpful to the coincidence problem. 

Instead, we then consider the case that only the field itself is coupled to the disformal metric. 
In the simplest case, we have the Einstein-Hilbert action for gravity coupled minimally to the matter sector and 
a canonical scalar field coupled to the barred metric. This seems to be a minimal set-up employing the relation (\ref{disf}), in 
the sense that all other matter than $\phi$ is minimally coupled to standard general relativity. 
The only unusual feature is then an effective self-coupling of the field $\phi$. Explicitly, we write 
\beq \label{action}
S & = & \int d^4 x\Bigg[\sqrt{-g}\left(\frac{1}{2\kappa^2}R+\mathcal{L}_m\right)  \nonumber 
\\ 
& - &  
\sqrt{-\bar{g}}\left(\frac{1}{2}\bar{g}^{\alpha\beta}\phi_{,\alpha}\phi_{,\beta}+V(\phi)\right)\Bigg],
\eeq
where $\mathcal{L}_m$ is the Lagrangian density for matter, and $\kappa^2 = 8 \pi G$. In the following, we denote 
\be \label{i}
I \equiv g^{\alpha\beta}\phi_{,\alpha}\phi_{,\beta}.
\ee
The scalar field has a canonical energy momentum tensor in the disformal metric and is covariantly conserved with 
respect to this {\it barred} metric, $\bar{\nabla}^\mu \bar{T}_{\mu\nu}^{(\phi)} = 0$. However, the physical metric is 
$g_{\mu\nu}$, and to avoid confusion, we rather write everything in terms of this unbarred metric and its Levi-Civita connection. The 
field couples also, through the relation (\ref{disf}) to the physical metric. One may thus associate an effective 
energy momentum tensor with the field. It can be shown to have the form 
\beq \label{emt}
T_{\mu\nu}^{(\phi)} & = & \frac{1}{\sqrt{A(A+IB)}}\Bigg[ 
\left(\frac{2A+IB}{2(A+IB)}+BV\right)\phi_{,\mu}\phi_{,\nu} \nonumber \\ & - & 
\left(\frac{1}{2}I + (A+IB)V\right)g_{\mu\nu}\Bigg].
\eeq
This tensor is also covariantly conserved, but with respect to the {\it unbarred} metric, consistently with the generalized Bianchi identity 
\cite{Koivisto:2005yk}. It is easy to see that this energy-momentum tensor can be put in the perfect-fluid form 
$T_{\mu\nu}^{(\phi)} = \rho u_{\mu}u_{\nu} + p\left(g_{\mu\nu} + u_{\mu}u_{\nu}\right)$ where the four-velocity is
$u_\mu = \phi_{,\mu}/\sqrt{-I}$. Thus the field cannot generate anisotropic stress. The form (\ref{emt}) makes also 
transparent that the model, in general, cannot be reduced to k-essence \cite{ArmendarizPicon:2000ah}. However, when $B$ is zero, 
the model can be reduced to canonic scalar field theory by the redefinition $\phi \rightarrow \int d\phi/\sqrt{A}$ 
(with just the potential possibly changing it's form from the original). Therefore a scalar field in a purely conformal 
metric is not essentially new model.   

\section{Cosmology} 

We then consider homogeneous and isotropic solutions cosmology with the action (\ref{action}). Consider the flat 
Friedmann-Robertson-Walker (FRW) metric,
\be \label{frw}
ds^2 = -dt^2 + a^2(t)(dx^2+dy^2+dz^2).
\ee
The expressions for the energy density and the pressure of the scalar field, which can be deduced from 
(\ref{emt}), have then, consistently with (\ref{rho2}), the rather simple forms,
\be \label{rho}
\rho^{(\phi)} = 
\frac{1}{\sqrt{1-\frac{B}{A}\dot{\phi}^2}}\left(\frac{\dot{\phi}^2}{2(A-B\dot{\phi}^2)} 
+ V \right)
\ee
\be \label{p}
p^{(\phi)} = 
\sqrt{1-\frac{B}{A}\dot{\phi}^2}\left(\frac{\dot{\phi}^2}{2(A-B\dot{\phi}^2)}    
- V \right)
\ee
The Friedmann equation gives the Hubble rate
\be \label{h1}
H^2 = \frac{8 \pi G}{3}\left(\rho^{(m)}+\rho^{(\phi)}\right),
\ee
and it's derivative can be found from
\be \label{h2}
2\dot{H} + 3H^2 = -\frac{8 \pi G}{3}\left(p^{(m)}+p^{(\phi)}\right),
\ee
where $\rho^{(m)}$ and $p^{(m)}$ are the energy density and pressure of the minimally coupled matter.

The interpretation is that the scalar field $\phi$ now lives in the metric
\beq \label{bfrw}
d\bar{s}^2 & = & -(A(t)-B(t)\dot{\phi}^2)dt^2 \nonumber \\ & + & A(t) a^2(t)(dx^2+dy^2+dy^2).
\eeq
Now $A$ is just the usual conformal factor. Let us thus consider what happens when the disformal factor $B$ is 
significant. Evidently, in cosmology the function $B$ corresponds to a distortion of the lapse function for the 
barred metric. As $B$ grows $B \gg A$, the time felt by the field begins to elapse slower (as measured by our unbarred clocks). As things 
happen 
faster in the matter frame, where distances and time intervals are measured with respect to the ''physical'', 
unbarred metric, the disformal field $\phi$ begins to look frozen when $B$ is large enough. This can be seen 
from the formulas (\ref{rho},\ref{p}), where the kinetic terms are suppressed by the square of the ratio of 
peculiar times as measured using the different metrics. However, the feature will always disappear when the field 
is a constant, and is thus a purely dynamical effect requiring some rolling of the scalar field. 

This suggests the following cosmological application. The cosmon field is well known to have so called tracking 
property, which guarantees it exhibits a constant ratio of the total energy density of the universe regardless of the initial 
conditions \cite{Wetterich:1994bg}. Hence, one may have a scalar field present during the whole evolution of the 
universe without fine-tuning. However, the field cannot explain the observed acceleration if it stays on the attractor. 
To toss the field off the attractor, one may reshape the potential \cite{Albrecht:1999rm,Copeland:2000vh}. A theoretically 
motivated possibility is to couple the field nonminimally to matter \cite{Amendola:1999er}, but 
this is known to lead to an instability at the linear level \cite{Koivisto:2005nr} (see also 
\cite{Valiviita:2008iv,Bean:2008ac} but however \cite{Bjaelde:2007ki,Corasaniti:2008kx}) and 
possible problems with quantum 
loop corrections \cite{Doran:2002bc}. However, considering neutrino coupling allows to link the acceleration scale with the
neutrino mass \cite{Amendola:2007yx}.  
Mechanisms based on nonminimal gravity couplings 
are employed in extended quintessence \cite{Perrotta:1999am}, Gauss-Bonnet quintessence \cite{Koivisto:2006xf} and 
in more general models with nonlinear functions of curvature invariants \cite{Nojiri:2004bi,Nojiri:2007te}. 

\begin{figure}
\begin{center}
\includegraphics[width=0.49\textwidth]{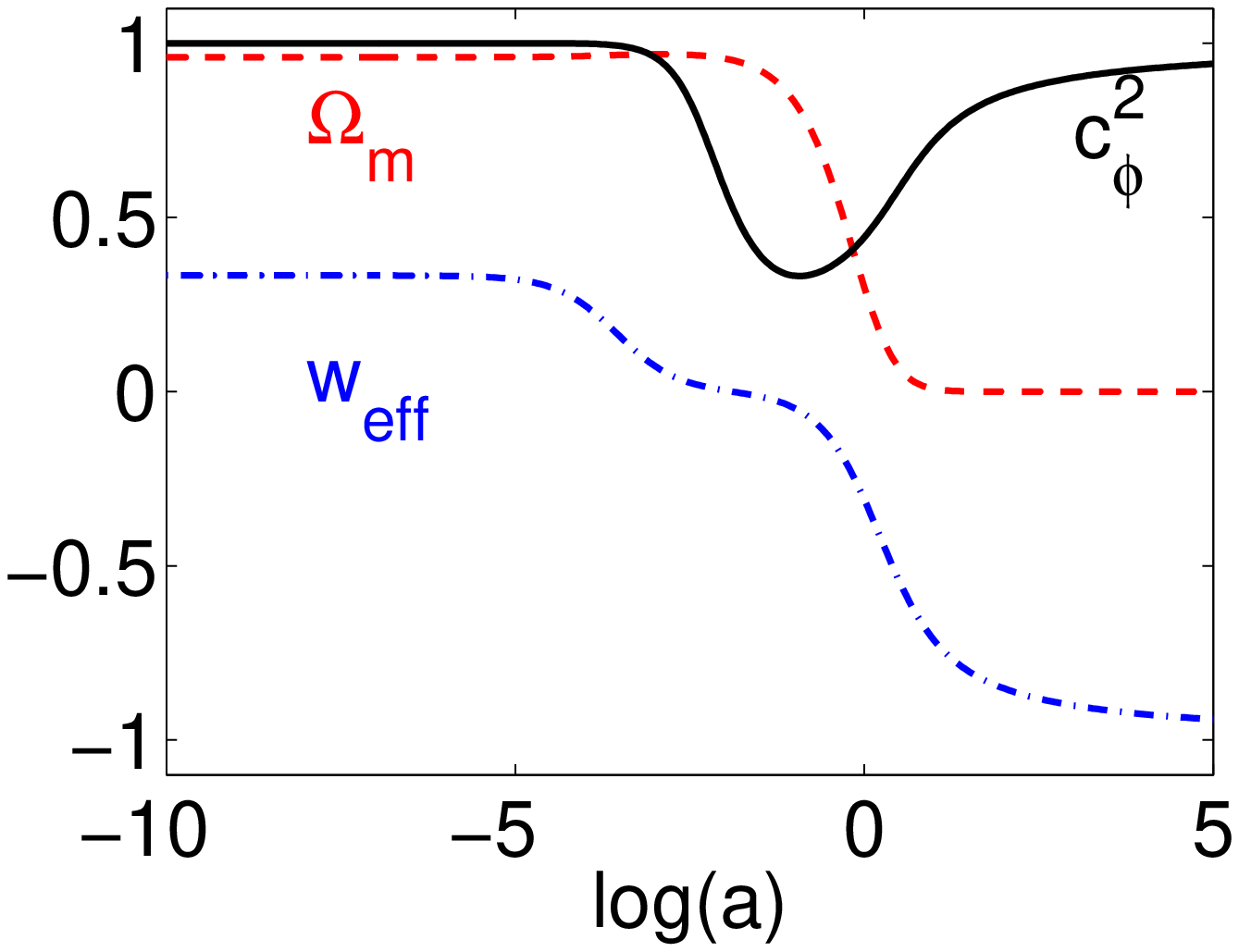}
\includegraphics[width=0.49\textwidth]{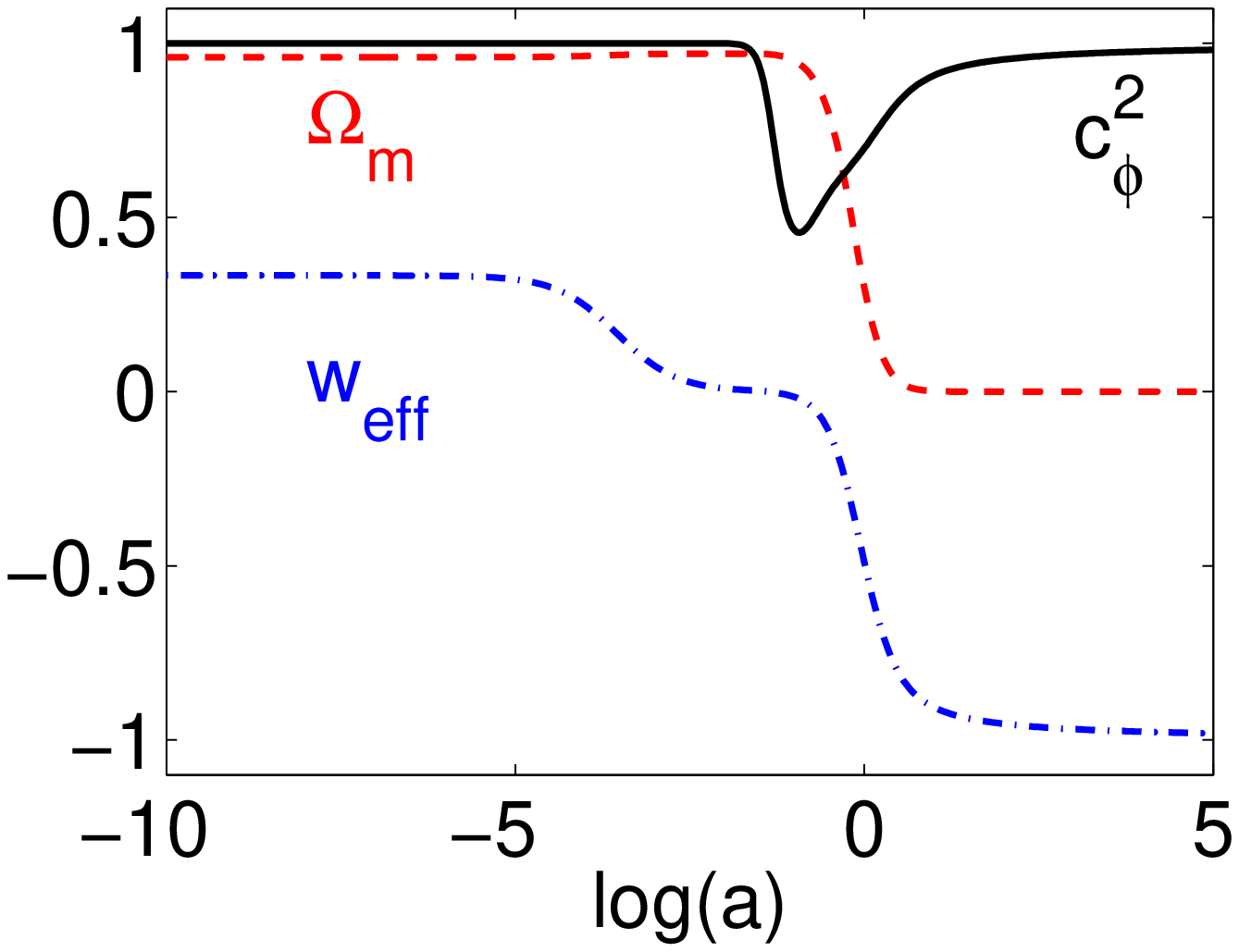}
\caption{\label{evol} Evolution of the model (\ref{model}) with two examples having $\Omega_m=0.3$ today. In 
the upper panel we have $\beta=15$ and in the lower panel $\beta=25$. The blue dash-dotted lines show the 
effective equation of state of the universe, and the red dashed lines show the relative amount of matter 
$\Omega_m$ as a function of the logarithm of the scale factor. The black solid line is the square of the 
sound speed of perturbations of the scalar field. We used $\lambda=10$ for the potential and for the 
radiation density $\Omega_r \approx 8.5 \cdot 10^{-5}.$}
\end{center}
\end{figure}

Here we instead let the disformal effect freeze the scalar field. Thus we consider the possibility 
that the lapse distortion $B$ redshifts the kinetic energy away, thus stopping the field and triggering a potential dominated era. Clearly, 
the potential of the field then provides an effective cosmological constant and the universe evolves into future de Sitter stage.
To look into this in more detail, let us specialize to the simple case
\be \label{model}
A=1, \quad B = B_0 e^{\beta\kappa\phi}, \quad V = V_0e^{\lambda\kappa\phi}.    
\ee
We set the purely conformal factor to unity to focus on the novel features. We choose the exponential forms for the 
disformal factor and the potential motivated by of high-energy physics considerations and convenience. Now, 
assuming initially $B \ll 1$, the tracking stage is characterized by 
\be
\kappa\phi = \frac{3(1+w_m)}{\lambda} + constant. 
\ee
Then the field mimics the background component with the equation of state $w_m$. By Eq.(\ref{h1}), the Hubble parameter 
drops as $ H^2 \sim a^{-3(1+w_m)} $. The disformal factor grows as $B \sim a^{3(1+w_m)\beta/\lambda}$. Therefore, 
the importance of the term $ B\dot{\phi}^2 $ is growing, and the tracking regime will eventually be interrupted 
iff $\beta > \lambda$. A similar condition exists for Gauss-Bonnet dark energy proposed in Ref. \cite{Nojiri:2005vv}: 
iff the slope of the Gauss-Bonnet coupling is steeper than that of the potential, dark energy domination occurs 
\cite{Koivisto:2006ai}. Numerical examples of the evolution in the present models are shown in Figure \ref{evol}.
There, as usual $\Omega_m = \kappa^2\rho_m/(3H^2)$ is the fractional matter density, and $w_{eff}$ is the effective 
pressure per density of the total matter content. The quantity $c_\phi^2$ is discussed in the next section.

\section{Fluctuations in the field}

The propagation of the perturbations of the field is characterized by the sound speed squared, $c_\phi^2=\delta 
p/\delta\rho|_{T_{oi}=0}$ \cite{Kodama:1985bj}. This is an important determinant of the physical properties of the effective fluid; 
one requires $c_\phi^2<1$ in order to eliminate the possibility of superluminal information exhange via
excitations of the field, and in addition one sees that if $c_\phi^2 < 0$ the perturbations are unstable and 
may blow up too fast. The sound speed by definition depends on the background. Here we consider the
FRW background as it includes Minkowski and de Sitter as particular limits. The sound speed is evaluated as the ratio
of pressure and density perturbation in the comoving frame, and therefore no gauge ambiguity arises. Consider
the so called total matter gauge, where the metric perturbations are parameterised by two longitudinal scalar 
potentials and the matter velocity vanishes. As we argued above, even the nonstandard scalar field does not generate
anisotropic stresses, and its velocity field is proportional to the field perturbation. It follows that the mentioned
longitudinal metric potentials are equal and the field $\phi$ is smooth in this frame. With these observations,
it becomes straightforward to obtain from (\ref{emt}) that 
\be \label{sound}
c_\phi^2 = \frac {\left(-\dot{\phi}^{2}B + 2ABV - 2\dot{\phi}^2 B^2V 
+2A\right)\left(1-\dot{\phi}^2\frac{B}{A}\right)}
{\dot{\phi}^2B + 2ABV-2\dot{\phi}^2B^2V+2A}.
\ee
One notes that when $B$ vanishes the sound speed squared becomes identically unity. It is a well known property of canonic 
fields that their perturbations always propagate with the speed of light. When $B$ is not identically zero, one gets nontrivial 
constraints on the model by requiring causality and stability.

In the scenario (\ref{model}) the sound speed deviates from unity at the present and during the matter dominated epoch 
when the disformal effect is freezing the field. The sound speed decreases to a smaller positive value and then evolves to back to unity. Two 
examples of this evolution are shown in Figure \ref{evol}. This confirms that the model is free of 
instabilities and that causality violations do not occur.
Compared to standard quintessence, one expects the disformal field to cluster somewhat more. The effective sound speed dips during the
critical era of structure formation, and thus the inhomogeneities forming in the field are dissipated less efficiently.
The possibility to exploit this feature to observationally distinguish the disformal from standard quintessence 
is left to future studies.   

To end, let us remark that flipping the sign of $B$ in this model
would make the sound speed of the field tachyonic. In fact, rather than acceleration, a singularity would 
occur. This is of the type II in the classification of Ref. \cite{Nojiri:2005sx}. The sudden future singularity 
is of the generalized, higher order nature \cite{Barrow:2004hk} in the sense that the second derivative of the 
Hubble rate diverges. Recently similar finite-time singularities have been found to possibly occur in $f(R)$ gravity 
models \cite{Nojiri:2008fk,Bamba:2008ut} and the nonlocally corrected gravity \cite{Koivisto:2008xfa, Koivisto:2008dh} proposed in 
\cite{Deser:2007jk,Nojiri:2007uq}. However, we consider here only the nonsingular case of positive $B$.

\section{Conclusion}

Modified gravity could unify inflation and the dark sector in cosmology. The standard frameworks for this
pursuit, the so called nonlinear gravity as a particular class of scalar-tensor theories, and quintessence models 
are connected by conformal transformation. In this article we investigated the possibility to use the disformal 
generalisation (\ref{disf}) as the fundamental relation between the two metrics. For the purpose of constructing 
a minimal set-up where this is possible, we considered the case where only the scalar field itself lives in the disformally 
related metric. Then all usual matter respects the equivalence principle. The disformal effect, a new type of effective 
self-interaction, is only seen when the field is dynamical. 

This can have interesting consequences for cosmological, rolling scalar fields. In particular, the disformal 
modification can freeze a tracking quintessence field in such a way that an accelerating de Sitter era follows 
a scaling matter era. For an exponential potential with the slope $\lambda$ and an exponential disformal factor with the 
slope $\beta$ the necessary condition for acceleration becomes very simply $\beta>\lambda$. The relative scale of the
potential and the disformal factor is then determined by requiring the observed relative abundance of matter today. This
mechanism resembles the scenario where the string-motivated exponential Gauss-Bonnet coupling of the Nojiri-Odintsov-Sasaki 
modulus triggers the acceleration.  

To determine the causality and stability of the propagating perturbations in the field, 
we derived an expression for the sound speed squared (\ref{sound}). 
Requiring consistency limits the possible solutions. These nontrivial conditions are generically satisfied by the
scenarios described above. This seems, together with the previous considerations of gravitational models of inflation and dark matter mentioned in the 
introduction, to suggest that a subtle modification of the relation of matter and gravitational 
metrics through the disformal transformation could also have a role in the resolution of present cosmological problems. 

\bibliography{refs}

\end{document}